\documentclass[twocolumn,showpacs,preprintnumbers,ams
m a t h , a m s symb]{revtex4} 
\usepackage[dvips]{graphicx}
\usepackage{verbatim}
\begin{document}

\preprint{x}

\author{A C Graham}
\affiliation{Cavendish Laboratory, Madingley Road,
Cambridge, CB3 OHE, United Kingdom}

\author{D L Sawkey}
\affiliation{Cavendish Laboratory, Madingley Road,
Cambridge, CB3 OHE, United Kingdom}

\author{M Pepper}
\affiliation{Cavendish Laboratory, Madingley Road,
Cambridge, CB3 OHE, United Kingdom}

\author{M Y Simmons$^{\ast}$}
\affiliation{Cavendish Laboratory, Madingley Road,
Cambridge, CB3 OHE, United Kingdom}

\author{D A Ritchie}
\affiliation{Cavendish Laboratory, Madingley Road,
Cambridge, CB3 OHE, United Kingdom}

\title{Energy-level pinning and the 0.7 spin state in one dimension:  GaAs quantum wires studied using finite-bias spectroscopy}

\date{\today}

\begin{abstract}
We study the effects of electron-electron interactions on the energy levels of GaAs quantum wires using finite-bias spectroscopy. We probe the energy spectrum at zero magnetic field, and at crossings of opposite-spin-levels in high in-plane magnetic field, $B$. Our results constitute direct evidence that spin-up (higher energy) levels pin to the chemical potential, $\mu$, as they populate. We also show that spin-up and spin-down levels abruptly rearrange at the crossing in a manner resembling the magnetic phase transitions predicted to occur at crossings of Landau levels. This rearranging and pinning of subbands provides a phenomenological explanation for the 0.7 structure, a 1D nanomagnetic state, and its high-$B$ variants.
\end{abstract}

\maketitle

\section{INTRODUCTION}
Whereas many collective electron phenomena in two- and zero-dimensional electron systems are well understood, such as the fractional quantum Hall effect and Kondo effect, this cannot be said of one-dimensional electron systems (1DES). As these may conceivably form the building-blocks of quantum circuits, it is important that their properties are understood. Theoretically, an interacting 1DES can be treated as a Luttinger-Liquid (LL) \cite{haldane}; although tunnelling experiments in parallel semiconductor QWs \cite{auslaender} and carbon nanotubes \cite{lee} have shown evidence of Luttinger liquid behaviour, many QW characteristics cannot at present be understood within the Luttinger liquid framework.  In particular, a spin-related phenomenon known as the \textit{0.7 structure} \cite{thomas96,birdscience,fitzgerald} has long resisted quantitative explanation.

According to non-interacting electron theories, the conductance of a semiconductor 1DES is quantized at $N(2e^2/h)$, where $N$ 1D modes lie below the Fermi energy.  In real systems however, an additional plateau occurs at around $0.7\times 2e^2/h$ --- the 0.7 structure. This deceptively simple feature has attracted much experimental \cite{thomas96,abiprl,thomas98,thomas02,kristensenprb,reillyprl,marcus,rokhinson,rolfs} and theoretical \cite{spivak,bruus,klironomos,meir,wang96,karl05,rejecmeir} interest, because its unusual magnetic field ($B$) and temperature ($T$) dependences \cite{thomas96,marcus,reillyprl,kristensenprb} imply that complex electron spin interactions strongly influence the behaviour of even the simplest quantum devices.

\begin{figure}[t!]
\begin{center}
\includegraphics[width=\columnwidth]{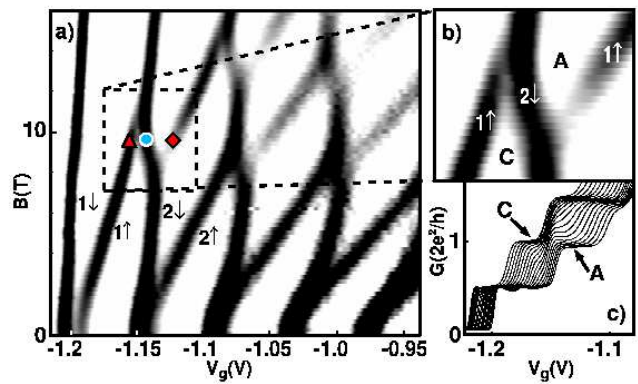}
\end{center}
\caption{(Color online) Evolution and crossings of 1D subbands in $B$. \textbf{(a)} Grey-scale diagram of $dG/dV_{\rm{\rm{g}}}$ as a function of $V_{\rm{\rm{g}}}$ for $B=0$ to $16~$T in increments of $0.2~$T. White represents conductance plateaux, and dark lines correspond to a subband populating. Right-moving (left-moving) lines are higher energy spin-up (lower energy spin-down) subbands. We refer to the three symbols later in the paper. \textbf{(b)} A close-up of the 1$\uparrow$ and 2$\downarrow$ crossing.  The trajectory of 1$\uparrow$ is discontinuous, with its two parts overlapping in $B$.  \textbf{A} and \textbf{C} indicate the non-quantized analog and complement structure respectively, shown in \textbf{(c)} Conductance traces from $5.8~$T (left) to $13~$T, covering the range of figure (b).  The analog \textbf{A} (a variant of the 0.7 structure) and the complement \textbf{C} are indicated. \label{Fig1}}
\end{figure}

The 0.7 structure evolves continuously into the lowest energy spin-down mode with increasing $B$, implying that it is a type of spontaneously spin-polarised state.  Whereas the 0.7 structure occurs at the $B=0$ crossing of the 1$\uparrow$ and 1$\downarrow$ subbands, related conductance structures called `analogs' have recently been discovered at the crossing of the 1$\uparrow$ and 2$\downarrow$ subbands in high in-plane $B$ \cite{abiprl}. In the region of the analogs, energy levels of opposite spin abruptly rearrange as they populate, forming a completely spin-polarised state. This is thought to be driven by the resulting exchange energy enhancement \cite{abissc,karl05} and resembles the magnetic phase transitions predicted to occur at crossings of Landau levels \cite{giuliani}. In this paper, we provide direct evidence from DC-bias spectroscopy \cite{glazman,patel91a}, that the 0.7 structure and analogs are caused by the highest energy spin-up subband pinning to the chemical potential, $\mu$, as predicted by Kristensen and Bruus \cite{kristensenprb}, together with an abrupt rearranging of spin-up and spin-down subbands.

\section{SAMPLES AND MEASUREMENT}
Our samples consist of split-gate devices defined by electron beam lithography on a Hall bar etched from a high mobility GaAs/Al$_{0.33}$Ga$_{1-0.33}$As heterostructure.  The two-dimensional electron gas lies $292~$nm below the surface of the heterostructure. All the 1DES samples used in this work have a lithographic length of 0.4~$\mu$m and a width of 0.7~$\mu$m. We used an in-plane $B$ aligned perpendicular to the current direction.  We have however observed the same behaviour for in-plane parallel $B$. By monitoring the Hall voltage, the out-of-plane misalignment was measured to be $0.3\,^{\circ}$. The measurement temperature was $50~$mK.

\section{TWO NON-QUANTIZED CONDUCTANCE STRUCTURES RELATE TO REARRANGING OF SUBBANDS AT CROSSINGS}
 
The rearranging of the 1$\uparrow$ and 2$\downarrow$ subbands in the crossing region is inferred from the data in fig.~\ref{Fig1}(a), which exhibits multiple crossings of spin-split 1D subbands. The data can be thought of as an energy diagram, where the black lines represent subbands. The close-up of the crossing of 1$\uparrow$ and 2$\downarrow$ in fig.~\ref{Fig1}(b) shows that the trajectory of 1$\uparrow$ is discontinuous at the crossing with the two parts of 1$\uparrow$ overlapping in $B$.  I.e., at around $B=10~$T, going from left to right in gate-voltage, $V_{\rm{g}}$, 1$\uparrow$ populates \textit{twice}, at two different $V_{\rm{g}}$.  Thus, the 1DES energy spectrum is not fixed, but rearranges as the subbands populate, an effect thought to be due to e-e interactions \cite{abissc,karl05,giuliani}. 

Two plateau regions, \textbf{A} and \textbf{C} in fig.~\ref{Fig1}(b), are formed between the overlapping parts of 1$\uparrow$, on the right and left of 2$\downarrow$. Although \textbf{A} and \textbf{C} are non-quantized conductance structures (fig.~\ref{Fig1}(c)), they are separated by a constant quantized conductance of $\sim0.5(2e^2/h)$. \textbf{A}, the \textit{0.7 analog}, has similar properties to the 0.7 structure \cite{abiprl,thomas96}; in fig.~\ref{Fig1}(b) the analog region \textbf{A} at $B=10~$T and above is equivalent to the 0.7 structure region near $B=0$. However, the region below $10~$T has no equivalent at $B=0$ --- one cannot investigate $|B|<0$.  In this region (fig.~\ref{Fig1}(b)), we find a new non-quantized feature, \textbf{C}, called a \textit{complement} structure (see fig.~\ref{Fig1}(c)).  As we will show, the DC-bias characteristics of the complement, analog and 0.7 structure provide evidence that they are all caused by pinning of a spin-up subband together with an abrupt drop in energy of a spin-down subband.

 \begin{figure}
\begin{center}
\includegraphics[width=\columnwidth]{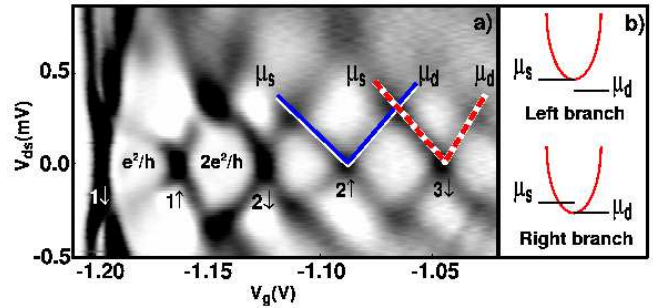}
\end{center}
\caption{(Color online) \textbf{(a)} Grey-scale of $dG/dV_{\rm{\rm{g}}}$ data at $5~$T as
a function of DC-bias $V_{\rm{ds}}$ and $V_{\rm{\rm{g}}}$. Labels indicate whether a branch corresponds to a subband intercepting $\mu_{\rm{s}}$ or $\mu_{\rm{d}}$, as shown in \textbf{(b)}. A left (right) branch in $+V_{\rm{ds}}$ corresponds to a subband intercepting the source (drain) chemical potential $\mu_{\rm{s}}=\mu+eV_{\rm{ds}}/2$ ($\mu_{\rm{d}}=\mu-eV_{\rm{ds}}/2$).\label{Fig2}}
\end{figure}

\section{BIAS SPECTROSCOPY OF CROSSINGS OF SPIN SUBBANDS IN HIGH MAGNETIC FIELDS}

DC-bias ($V_{\rm{ds}}$) data taken at $5~$T (fig.~\ref{Fig2}(a)) gives insight into $V_{\rm{ds}}$ characteristics at the crossing and $B=0$; the $5~$T data is simpler than these regimes, because the spin subbands are far apart in energy. At $V_{\rm{ds}}=0$ in fig.~\ref{Fig2}(a), each subband gives one dark feature as it intercepts $\mu$. Each of these features splits into a V-shaped pair of branches at $V_{\rm{ds}}>0$ because $\mu$ splits in two, $\mu_{\rm{s}}$ ($\mu_{\rm{d}}$) for the source (drain). Left (right) branches are due to subbands intercepting $\mu_{\rm{s}}$ ($\mu_{\rm{d}}$) - see fig.~\ref{Fig2}(b). The $+V_{\rm{ds}}$ branches associated with 2$\uparrow$ are marked with a solid blue V-shape and the branches associated with 3$\downarrow$ are marked with a dashed red V-shape (n.b. here, and below, the term `V-shape' refers to two branch features in the $+V_{\rm{ds}}$ part of the data).  In this $5~$T data, spin-up features in particular are generally consistent with the non-interacting $V_{\rm{ds}}$ model \cite{glazman} in which \textit{both branches} of the V-shape are present at any $B$ because the subband must pass through both $\mu_{\rm{s}}$ and $\mu_{\rm{d}}$. 

\begin{figure*}
\begin{center}
\includegraphics[width=\textwidth]{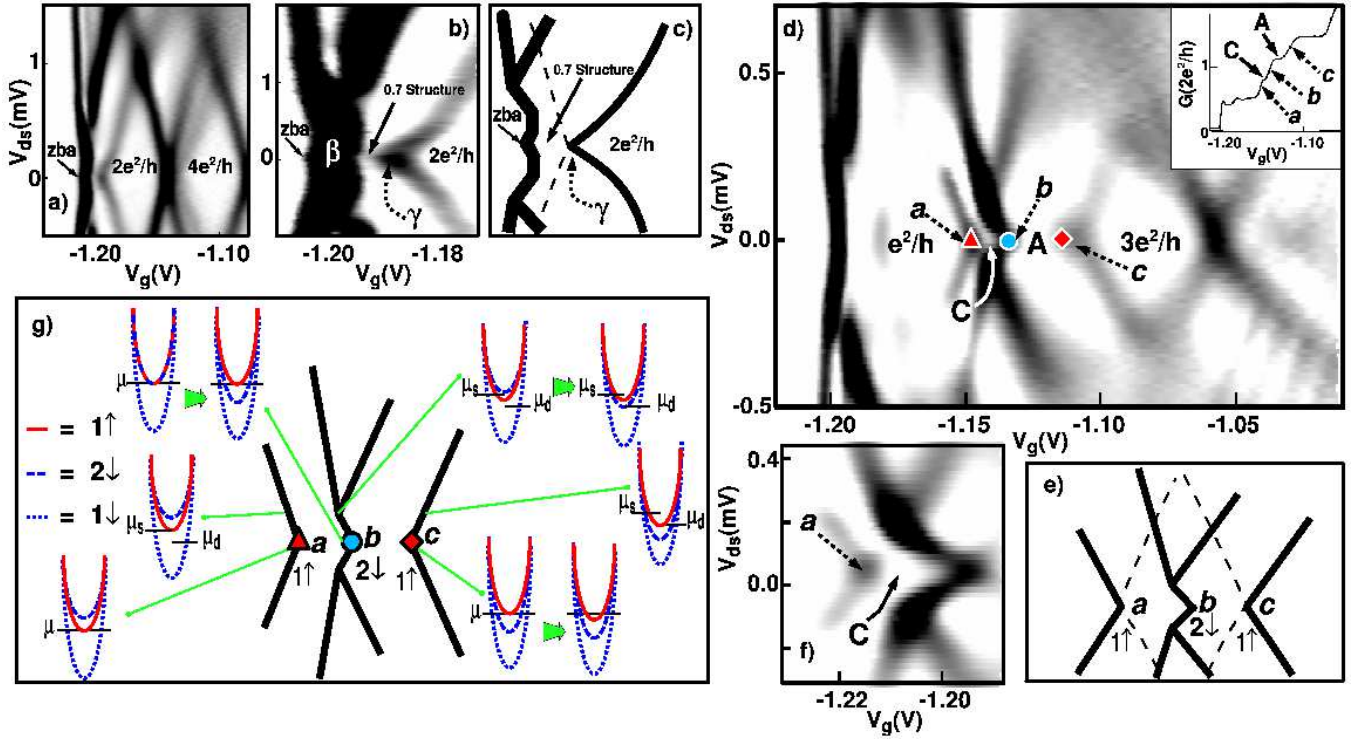}
\end{center}
\caption{(Color online) \textbf{(a)} Grey-scale of $dG/dV_{\rm{\rm{g}}}$ at $B=0$ as a
function of $V_{\rm{ds}}$.  White regions are plateaux. A close-up \textbf{(b)} shows that $\gamma$, separating the 0.7 structure from $2e^2/h$, does not split in $V_{\rm{ds}}$ - the left branch is absent. This is illustrated schematically in \textbf{(c)}, where the `missing left branches' are represented by dashed lines. \textbf{(d)} $V_{\rm{ds}}$ data at the crossing at $B=9~$T. At $V_{\rm{ds}}=0$, \textbf{\textit{a}}, \textbf{\textit{b}} and \textbf{\textit{c}} correspond to the 1$\uparrow$, 2$\downarrow$ and 1$\uparrow$ features marked by symbols in fig.~\ref{Fig1}. \textbf{\textit{a}} has no right branch in $V_{\rm{ds}}$ and \textbf{\textit{c}} has no left branch. \textbf{\textit{b}} and \textbf{\textit{c}} beside the analog \textbf{A} are equivalent to $\beta$ and $\gamma$ beside the 0.7 structure in (b). Inset:  Conductance trace for $V_{\rm{ds}}=0$ at $B=9~$T. \textbf{(e)} The `missing branches' for \textbf{\textit{a}} and \textbf{\textit{c}} are represented by dashed lines in this schematic diagram of the crossings. \textbf{(f)} A close-up of \textbf{\textit{a}} and \textbf{\textit{b}}, in data from a similar sample at $8.6~$T, demonstrates that \textbf{\textit{a}} has no right branch. \textbf{(g)} A schematic of (d), showing the configurations of subbands. Missing branches indicate that 1$\uparrow$ is pinned to $\mu$ in the complement and analog regions \textbf{A} and \textbf{C} (see text).  
\label{Fig3}}
\end{figure*}

In contrast, in the $V_{\rm{ds}}$ data at $B=0$ (figs.~\ref{Fig3}(a) and (b)), there are strong deviations from the expected non-interacting behaviour. Moving from $5~$T (fig.~\ref{Fig2}(a)) to $B=0$ (fig.~\ref{Fig3}(a) and (b)), the $e^2/h$ plateau evolves into the 0.7 structure.  The V-shaped 1$\uparrow$ feature which separates the $e^2/h$ plateau from the $2e^2/h$ plateau moves to the left, forming the grey line marked $\gamma$, that separates the 0.7 structure from the $2e^2/h$ plateau. However, $\gamma$ is no longer a V-shape as it has no left branch --- it is just a single right-moving branch (fig.~\ref{Fig3}(b) and (c)); the expected but `missing' left branch is represented by a dashed line in the schematic diagram in fig.~\ref{Fig3}(c). $\gamma$ relates to the 1$\uparrow$ subband, so at $B=0$ although we can detect 1$\uparrow$ intercepting $\mu_{\rm{d}}$ --- the right-moving branch from $\gamma$ --- the branch indicating that 1$\uparrow$ has intercepted $\mu_{\rm{s}}$ is missing. We will show that this unexpected behaviour is direct evidence that the 0.7 structure is caused by pinning of the 1$\uparrow$ subband as it populates.

Branches are also missing in the $V_{\rm{ds}}$ data at the crossing (fig.~\ref{Fig3}(d)), in the region of the analog and complement --- the inset to fig.~\ref{Fig3}(d) shows conductance at the crossing for $V_{\rm{ds}}=0$.  Fig.~\ref{Fig3}(e) gives a schematic of the main features of the crossing \footnote{Between taking the fig.~\ref{Fig1} data and the figs.~\ref{Fig2} and \ref{Fig3} data, the pinch-off voltage changed by $\sim0.07$V. We have added $0.07~$V to $V_{\rm{g}}$ in fig.~\ref{Fig1} to aid comparison with the other figures.  The change in the device also caused the crossing to shift by $-0.5~$T, so the features in fig.~\ref{Fig2} compare better to $B=9.5~$T in fig.~\ref{Fig1}, rather than $9~$T.}. Again, compare fig.~\ref{Fig3}(d) to the $5~$T data in fig.~\ref{Fig2}(a); the 1$\uparrow$ V-shape at $5~$T which separates the $e^2/h$ and $2e^2/h$ plateaux has at $9~$T shifted to the right to form features \textbf{\textit{a}} \textit{and} \textbf{\textit{c}} in fig.~\ref{Fig3}(d),  whilst 2$\downarrow$ causes feature \textbf{\textit{b}}.  At $V_{\rm{ds}}=0$, points marked with coloured symbols correspond to the symbols in fig.~\ref{Fig1}(a), and the alternation between 1$\uparrow$ (feature \textbf{\textit{a}}), 2$\downarrow$ (feature \textbf{\textit{b}}) and 1$\uparrow$ (feature \textbf{\textit{c}}) again, indicates that 1$\uparrow$ and 2$\downarrow$ rearrange as they populate. Feature \textbf{\textit{a}} at the left edge of the complement structure \textbf{C} has no right branch in $V_{\rm{ds}}$ --- see the closeup in fig.~\ref{Fig3}(f), and the schematic in fig.~\ref{Fig3}(e) in which the absent right branch is represented with a dashed line. Feature \textbf{\textit{c}}, on the right of the analog \textbf{A}, has no left branch, and is equivalent to $\gamma$ in the region of the 0.7 structure in fig.~\ref{Fig3}(b); feature \textbf{\textit{b}} is equivalent to $\beta$. In short, whereas the spin-down feature \textbf{\textit{b}} splits into two branches with increasing $V_{\rm{ds}}$, the spin-up features \textbf{\textit{a}} and \textbf{\textit{c}} do not split and only have one branch, either right- or left-moving in $V_{\rm{ds}}$.  The absence of these branches cannot be understood in a non-interacting electron picture.

\subsection{Missing branches in the bias spectroscopy data indicates `pinning' of spin-up subbands}

The missing branches can be explained by a combination of two mechanisms --- the abrupt rearranging of the spin-up and spin-down subbands, together with simultaneous pinning of the spin-up subband to $\mu$. In figs.~\ref{Fig3}(d), (e) and (g), in finite $V_{\rm{ds}}$, 1$\uparrow$ intercepts $\mu_{\rm{s}}$ at the left-moving branch of feature \textbf{\textit{a}}, as illustrated by the schematic diagrams of subbands in fig.~\ref{Fig3}(g). At $V_{\rm{ds}}=0$, $\mu_{\rm{s}}$ is the same as $\mu$, so 1$\uparrow$ intercepts $\mu$ at \textbf{\textit{a}}. In finite $V_{\rm{ds}}$, 1$\uparrow$ intercepts $\mu_{\rm{d}}$ at the right-moving branch from feature \textbf{\textit{c}}. At $V_{\rm{ds}}=0$, $\mu_{\rm{d}}$ is the same as $\mu$, so 1$\uparrow$ \textit{must still be at} $\mu$ at feature \textbf{\textit{c}} --- 1$\uparrow$ reaches $\mu$ at \textbf{\textit{a}}, and remains close to $\mu$ until feature \textbf{\textit{c}}. In other words, for the left and right branches of 1$\uparrow$ (\textbf{\textit{a}} and \textbf{\textit{c}}) to be separated by such a large range in $V_{\rm{g}}$, then at $V_{\rm{ds}}=0$, 1$\uparrow$ must pin to $\mu$ from point \textbf{\textit{a}} until point \textbf{\textit{c}} throughout the regions of the analog and complement.   Thus, missing branches on the V-shaped features in the DC-bias data lead directly to the conclusion that spin-up subbands pin close to the chemical potential over a range of gate-voltages.
 
\subsection{The contrasting behaviour of spin-up and spin-down subbands, and their rearranging in energy at the crossings}
Unlike the 1$\uparrow$ subband, 2$\downarrow$ does not pin to $\mu$. Also using DC-bias spectroscopy, we have found that spin-down subbands do not give a simple V-shaped feature in $V_{\rm{ds}}$ \cite{abiprb}. The form of feature \textbf{\textit{b}} for 2$\downarrow$ in fig.~\ref{Fig3}(d), (e) and (g) is typical of spin-down subbands in general. The two branches of the V are not individually resolvable until a certain $V_{\rm{ds}}$, here $0.1~$mV, has been reached --- it is as if the expected V-shaped feature has been `collapsed' along the $V_g$ axis, so the left and right-branches lie on top of each other until $V_{\rm{ds}}=0.1~$mV. This implies the exact opposite behaviour for spin-down subbands than for spin-up --- it implies that spin-down subbands populate very abruptly, passing through both $\mu_{\rm{s}}$ and $\mu_{\rm{d}}$ within a very narrow gate-voltage range, even when $\mu_{\rm{s}}$ and $\mu_{\rm{d}}$ are separated in energy by more than $0.1~$mV, and in some cases, as much as $0.5~$mV \cite{abiprb}. In contrast, for spin-up subbands, it is as if the expected V-shaped feature has been `stretched' along the $V_g$ axis, so the left and right-branches lie far apart from each other in gate-voltage, indicating that spin-up subbands populate very gradually.  

Since 2$\downarrow$ populates abruptly at \textbf{\textit{b}}, 1$\uparrow$ and 2$\downarrow$ also rearrange in energy between the complement \textbf{C} and analog \textbf{A} regions, but with 1$\uparrow$ remaining pinned throughout. This rearranging resembles the exchange-driven magnetic phase-transitions predicted for Landau-level crossings \cite{giuliani,abissc,karl05}, and the combination of pinning of spin-up subbands with a sudden drop in energy of spin-down subbands provides an explanation \cite{abiprb} for why the 0.7 structure and analogs, and spin-down features in general \cite{abipe}, remain visible at surprisingly high temperatures.

\section{PINNING OF SPIN-UP SUBBANDS IS THE PHENOMENOLOGICAL ORIGIN OF THE NON-QUANTIZED 0.7, COMPLEMENT AND ANALOG STRUCTURES}

\begin{figure}[h!]
\begin{center}
\includegraphics[width=\columnwidth]{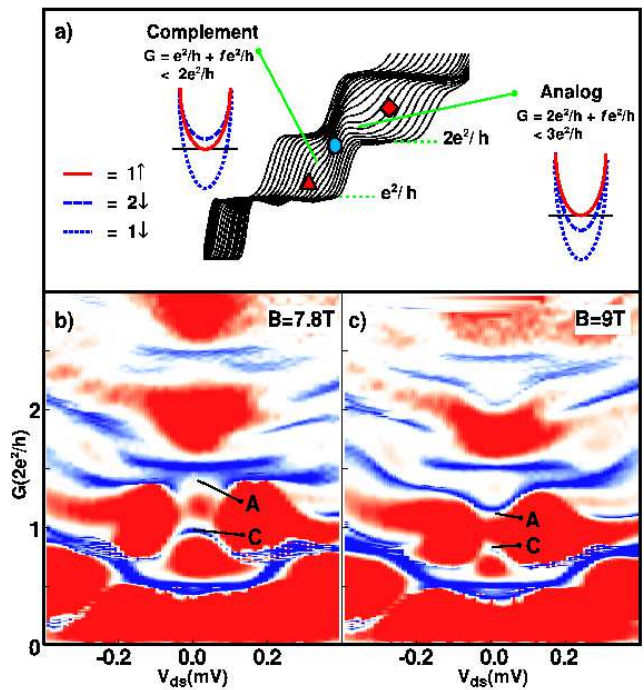}
\end{center}
\caption{(Color) \textbf{(a)} Conductance traces from fig.~\ref{Fig1}(a) with schematics illustrating how pinning of 1$\uparrow$ explains the complement and analog. Again, symbols indicate rises in conductance that correspond to features in figs.~\ref{Fig1} and \ref{Fig3}(d) and (g). \textbf{(b)} $dG/dV_{\rm{\rm{g}}}$ as a function of $G$ and $V_{\rm{ds}}$ at the crossing at $B=7.8~$T. Blue indicates plateaux, red indicates abrupt changes in $G$ with $V_{\rm{g}}$ and white indicates slowly changing $G$. \textbf{(c)} Similar data for $B=9~$T (also the data used to make fig.~\ref{Fig3} (d)). From $7.8~$T to $9~$T, the analog, \textbf{A}, strengthens as $G_{\rm{analog}}$ decreases, and the complement, \textbf{C}, weakens as $G_{\rm{complement}}$ decreases. $G_{\rm{complement}}$ and $G_{\rm{analog}}$ change immediately in finite $V_{\rm{ds}}$, unlike the quantized $1.5(2e^2/h)$ plateau which remains at fixed $G$ until it disappears at $V_{\rm{ds}}\sim \pm 0.2~$mV.\label{Fig4}}
\end{figure}

Pinning of 1$\uparrow$ also explains the non-quantized conductances of the complement and analog (fig~\ref{Fig4}(a)).  At $T>0$, a subband, ${N\sigma}$, lying close to $\mu$ gives a conductance of less than
$e^2/h$, because (ignoring tunnelling and reflection) $G_{N\sigma}=G_{0}f({\Delta}E,T)$ where $G_0=e^2/h$ and $f$ is the Fermi function, $T$ is
temperature and ${\Delta}E$ is the energy difference between $\mu$
and the bottom of the ${N\sigma}$ subband. If 1$\uparrow$ populates only
partially at the complement structure and pins close to $\mu$ over a range of gate-voltages, then the conductance of this subband, $G_{\rm{1\uparrow}}$, will be non-quantized and less than $e^2/h$,  i.e., the total conductance of the complement structure $G_{\rm{complement}}=G_{1\downarrow}+G_{1\uparrow}=e^2/h+fe^2/h<2e^2/h$. In earlier work \cite{abiprb}, we demonstrated that in contrast to spin-up subbands, spin-down subbands
populate very abruptly and do not pin to $\mu$. We also know from fig.~\ref{Fig1}(a) that 2$\downarrow$ populates between the complement and analog structures. Thus, a quantized
increase of $G_{2\downarrow} \sim e^2/h$ is expected when 2$\downarrow$ populates, and  $G_{\rm{analog}}=G_{1\downarrow}+G_{2\downarrow}+G_{1\uparrow}=e^2/h+e^2/h+fe^2/h<3e^2/h$, i.e., the analog conductance is $\sim e^2/h$ greater than the complement conductance throughout the crossing region, because of the population of 2$\downarrow$ between the two structures.  Above the analog, 1$\uparrow$ goes from being partially to fully populated, giving an increase in $G$ of less than $e^2/h$, and a total quantized conductance of $3e^2/h$.

We can perform similar analysis for the 0.7 structure at $B=0$ by considering 1$\downarrow$ and 1$\uparrow$ instead of 2$\downarrow$ and 1$\uparrow$. Near the crossing, 2$\downarrow$ and 1$\uparrow$ cannot be degenerate when they first populate because we can resolve both features \textbf{\textit{a}} and \textbf{\textit{b}}. However, fig.~\ref{Fig3}(b) at $B=0$ has no equivalent to feature \textbf{\textit{a}}, thus 1$\downarrow$ and 1$\uparrow$ are degenerate when they first populate. Just as the subbands at the crossing rearrange in energy abruptly, at $B=0$, the degeneracy between 1$\downarrow$ and 1$\uparrow$ is abruptly lifted as they populate --- 1$\downarrow$ drops suddenly in energy  \cite{abiprb} to give $\beta$, whilst 1$\uparrow$ pins to $\mu$ between $\beta$ and $\gamma$, giving non-quantized conductance (as first proposed by Kristensen \textit{et al.} \cite{kristensenprb}), before populating fully at $\gamma$.  Below the onset of 
V-shaped splitting from $\beta$ (for $V_{\rm{ds}}<0.3~$mV), there is no left branch from $\gamma$ because 1$\downarrow$ and 1$\uparrow$ pass through $\mu_{\rm{s}}$ together; the missing left branch from 1$\uparrow$ is part of feature $\beta$, and is separated from the 1$\uparrow$ right branch at $\gamma$ by finite $V_{\rm{g}}$ because the subband is pinned.

$G_{\rm{complement}}$ and $G_{\rm{analog}}$ immediately change with $V_{\rm{ds}}$ (fig.~\ref{Fig4}(a), (b) and (c))\footnote{A linear correction was added to the data in fig.~\ref{Fig4}(b) and (c) to allow better resolution of features at both high and low $G$.}.  The analog \textbf{A} at $B=9~$T rises with increasing $V_{\rm{ds}}$ from $1.15(2e^2/h)$ to $\sim1.35(2e^2/h)$, just as the 0.7 structure rises to $\sim0.85(2e^2/h)$ in finite bias (see refs. \cite{patel91a,thomasPM}), whereas $G_{\rm{complement}}$ \textit{decreases} with increasing $V_{\rm{ds}}$. This is in stark contrast to the behaviour of quantized plateaux: for example, the $3e^2/h$ plateau in fig.~\ref{Fig4}(c) remains at the same $G$ with increasing $V_{\rm{ds}}$ until it disappears.  Quantized plateaux do not change conductance in $V_{\rm{ds}}$ because, by definition, they occur when the subband edge is some way below $\mu$. Therefore for moderate $V_{\rm{ds}}$, the subband still lies well below $\mu_{\rm{s}}$ and $\mu_{\rm{d}}$ and $G$ will be unaffected by the energy gap between $\mu_{\rm{s}}$ and $\mu_{\rm{d}}$.  The change in $G_{\rm{complement}}$ and $G_{\rm{analog}}$ at small $V_{\rm{ds}}$ is consistent with 1$\uparrow$ pinning close to $\mu$ at those features.

\section{DISCUSSION AND CONCLUSIONS}
It has been proposed that the rise in $G$ of the 0.7 structure with decreasing $T$ may relate to the Kondo effect \cite{marcus}. The basis for this theory is the `zero-bias anomaly' (ZBA), a peak in $G$ at $V_{\rm{ds}}=0$, similar to that observed in quantum dots \cite{goldhaberN}.  We routinely observe such ZBAs at $B=0$, which, in a greyscale diagram such as fig.~\ref{Fig3}(b), take the form of a narrow pointed feature in the pinch-off voltage at $V_{ds}=0$ (marked \textbf{zba} in figs.~\ref{Fig3}(b) and (c) and indicated by an arrow).  However, we have not observed zero-bias anomalies in the conductance at the crossings, despite the presence of the analog structures. Analogs rise in conductance and disappear with decreasing $T$ --- if the disappearance of the analog and its large conductance enhancement at low $T$ were due to the Kondo effect, then the enhanced conductance would be destroyed by $V_{\rm{ds}}$, and hence, a ZBA would occur. The absence of any ZBA implies that Kondo physics is not the main cause of the enhanced $G$ associated with the 0.7 structure and analog variants at low $T$.

The phenomenological theory that spin-up subbands pin to $\mu$, but spin-down do not, provides a consistent interpretation for virtually all the characteristics of the 0.7 structure `family'. As previously observed \cite{kristensenprb}, pinning of a spin-up subband slightly below $\mu$ can explain why the 0.7 structure is typically absent at low $T$, but appears and decreases in $G$ as $T$ rises --- this also applies to the analogs at crossings. This is not, however, the only $T$ regime associated with the 0.7 structure.  In fact, the 0.7 structure only decreases in $G$ with increasing $T$ if it is above $\sim 0.6(2e^2/h)$ at low $T$ --- this depends on confining potential, which can be modified by applying a negative voltage to a `midline' gate \cite{thomas00}, or by using a scanning probe tip \cite{rolfs}.  It was observed that the 0.7 structure sits below $0.6(2e^2/h)$ for negative midline voltages, and certain scanning probe positions, and \textit{rises} in $G$ with increasing $T$. The same $T$ dependence is observed in an in-plane $B$ field --- once the 0.7 structure has moved below $\sim 0.6(2e^2/h)$ due to the $B$ field, it also rises in $G$ with increasing $T$ \cite{thomas02}. In other words, there is a crossover from one $T$ regime to the other, and a low $T$ conductance of $\sim 0.6(2e^2/h)$ marks the crossover. These two $T$ regimes also exist for the analog with a `crossover' conductance of $\sim 1.2(2e^2/h)$ in that case \cite{abiprl}. This second $T$ regime is also consistent with pinning and corresponds to the spin-up subband pinning slightly \textit{above} $\mu$. In addition, at the distinct crossover in $B$ between the two $T$ regimes, $G$ is invariant with $T$.  This is consistent with the subband pinning \textit{exactly at} $\mu$.  The rearranging of spin-up and spin-down subbands is also compatible with pinning of spin-up subbands.  Taken together, rearranging and pinning explain the presence of \textit{two} non-quantized structures in the crossing region (the complement and analog), their $V_{\rm{ds}}$ characteristics, and why these non-quantized structures are separated by a quantized conductance.

Additional evidence in support of our interpretation is that similar $V_{\rm{ds}}$ analysis applied to spin-down subbands explains why the 0.7 structure survives high temperatures \cite{abiprb}, and explains other spin-asymmetries \cite{abipe}.  Furthermore, pinning was suggested as an explanation for the unusual thermopower signature of the 0.7 structure \cite{appleyard}, and it is also consistent with the 0.7 structure shot-noise \cite{roche} signature. 
 
To conclude, we have used DC-bias spectroscopy to study the rearranging of spin-split subbands at crossings. Our results provide \textit{direct} evidence that spin-up subbands pin to $\mu$ in the region of the analog and complement structures, and the 0.7 structure. This, combined with the formation of a spin gap \cite{kristensenprb,reillyprl} and the abrupt drop in energy of the spin-down subband \cite{abiprb} explains the non-quantized conductances of these features, their temperature dependences, and the shot-noise \cite{roche} and thermopower \cite{appleyard} signatures of the 0.7 structure.  As yet, there is no theory that explains why spin-up subbands should pin in this way at crossings and $B=0$.  We hope that the evidence in this paper will provide the stimulus for theoretical work in this direction.

We thank K.-F. Berggren, N.R. Cooper, C. J. B. Ford, F. Sfigakis, V. Tripathi and K. J. Thomas for useful discussions. We also acknowledge the COLLECT European Research Training Network. This work was supported by EPSRC, UK. ACG acknowledges support from Emmanuel College, Cambridge.


\begin{thebibliography}{10}

\bibitem{haldane}
F.~D.~M. Haldane.
\newblock {\em J.\ Phys.\ C}, 14:2585, 1981.

\bibitem{auslaender}
O.~M. Auslaender, A.~Yacoby, R.~de~Picciotto, K.~W. Baldwin, L.~N. Pfeiffer,
  and K.~W. West.
\newblock {\em Science}, 295:825, 2002.

\bibitem{lee}
J.~Lee, S.~Eggert, H.~Kim, S.~J. Kahng, H.~Shinohara, and Y.~Kuk.
\newblock {\em Phys.\ Rev.\ Lett.}, 93:166403, 2004.

\bibitem{thomas96}
K.~J. Thomas, J.~T. Nicholls, M.~Y. Simmons, M.~Pepper, D.~R. Mace, and D.~A.
  Ritchie.
\newblock {\em Phys.\ Rev.\ Lett.}, 77:135, 1996.

\bibitem{birdscience}
J.~P. Bird and Y.~Ochiai.
\newblock {\em Science}, 303:1621, 2004.

\bibitem{fitzgerald}
R.~Fitzgerald.
\newblock {\em Phys. Today}, 55(5):21, 2002.

\bibitem{abiprl}
A.~C. Graham, K.~J. Thomas, M.~Pepper, N.~R. Cooper, M.~Y. Simmons, and D.~A.
  Ritchie.
\newblock {\em Phys.\ Rev.\ Lett.}, 91:136404, 2003.

\bibitem{thomas98}
K.~J. Thomas, J.~T. Nicholls, N.~J. Appleyard, M.~Y. Simmons, M.~Pepper, D.~R.
  Mace, W.~R. Tribe, and D.~A. Ritchie.
\newblock {\em Phys.\ Rev.\ B}, 58:4846, 1998.

\bibitem{thomas02}
K.~J. Thomas, J.~T. Nicholls, M.~Pepper, M.~Y. Simmons, D.~R. Mace, and D.~A.
  Ritchie.
\newblock {\em Physica E}, 12:708, 2002.

\bibitem{kristensenprb}
A.~Kristensen, H.~Bruus, A.~E. Hansen, J.~B. Jensen, P.~E. Lindelof, C.~J.
  Marckmann, J.~Nygard, C.~B. S{\"o}renson, F.~Beuscher, A.~Forchel, and
  M.~Michel.
\newblock {\em Phys.\ Rev.\ B}, 62:10950, 2000.

\bibitem{reillyprl}
D.~J. Reilly, T.~M. Buehler, J.~L. O'Brien, A.~R. Hamilton, A.~S. Dzurak, R.~G.
  Clark, B.~E. Kane, L.~N. Pfeiffer, and K.~W. West.
\newblock {\em Phys.\ Rev.\ Lett.}, 89:246801, 2002.

\bibitem{marcus}
S.~M. Cronenwett, H.~J. Lynch, D.~Goldhaber-Gordon, L.~P. Kouwenhoven, C.~M.
  Marcus, K.~Hirose, N.~S. Wingreen, and V.~Umansky.
\newblock {\em Phys.\ Rev.\ Lett.}, 88:226805, 2002.

\bibitem{rokhinson}
L.~P. Rokhinson, L.~N. Pfeiffer, and K.~W. West.
\newblock {\em Phys.\ Rev.\ Lett.}, 96:156602, 2006.

\bibitem{rolfs}
R.~Crook, J.~Prance, K.~J. Thomas, S.~J. Chorley, I.~Farrer, D.~A. Ritchie,
  M.~Pepper, and C.~G. Smith.
\newblock {\em Science}, 312:1359, 2006.

\bibitem{spivak}
B.~Spivak and F.~Zhou.
\newblock {\em Phys.\ Rev.\ B}, 61:16730, 2000.

\bibitem{bruus}
H.~Bruus, V.~V. Cheianov, and K.~Flensberg.
\newblock {\em Physica E}, 10:97, 2001.

\bibitem{klironomos}
A.~D. Klironomos, J.~S. Meyer, and K.~A. Matveev.
\newblock {\em Europhys.\ Lett.}, 74, 2006.

\bibitem{meir}
Y.~Meir, K.~Hirose, and N.~S. Wingreen.
\newblock {\em Phys.\ Rev.\ Lett.}, 89:196802, 2002.

\bibitem{wang96}
C.-K. Wang and K.-F. Berggren.
\newblock {\em Phys.\ Rev.\ B}, 54:14257, 1996.

\bibitem{karl05}
K.-F. Berggren, P.~Jaksch, and I.~I. Yakimenko.
\newblock {\em Phys.\ Rev.\ B}, 71:115303, 2005.

\bibitem{rejecmeir}
T.~Rejec and Y.~Meir.
\newblock {\em Nature}, 442:900, 2006.

\bibitem{abissc}
A.~C. Graham, K.~J. Thomas, M.~Pepper, M.~Y. Simmons, D.~A. Ritchie, K.-F.
  Berggren, P.~Jaksch, A.~Debnarova, and I.~I. Yakimenko.
\newblock {\em Sol.\ St.\ Commun.}, 131:591, 2004.

\bibitem{giuliani}
G.~F. Giuliani and J.~J. Quinn.
\newblock {\em Phys.\ Rev.\ B}, 31:6228, 1985.

\bibitem{glazman}
L.~I. Glazman and A.~V. Khaetskii.
\newblock {\em Europhys.\ Lett.}, 9:263, 1989.

\bibitem{patel91a}
N.~K. Patel, J.~T. Nicholls, L.~Mart\'{\i}n-Moreno, M.~Pepper, J.~E.~F. Frost,
  D.~A. Ritchie, and G.~A.~C. Jones.
\newblock {\em Phys.\ Rev.\ B}, 44:13549, 1991.

\bibitem{abiprb}
A.~C. Graham, M.~Pepper, M.~Y. Simmons, and D.~A. Ritchie.
\newblock {\em Phys.\ Rev.\ B}, 72:193305, 2005.

\bibitem{abipe}
A.~C. Graham, K.~J. Thomas, M.~Pepper, M.~Y. Simmons, and D.~A. Ritchie.
\newblock {\em Physica E}, 22:264, 2004.

\bibitem{thomasPM}
K.~J. Thomas, J.~T. Nicholls, M.~Y. Simmons, M.~Pepper, D.~R. Mace, and D.~A.
  Ritchie.
\newblock {\em Philos. Mag. B}, 77:1213, 1998.

\bibitem{goldhaberN}
D.~Goldhaber-Gordon, H.~Shtrikman, D.~Mahalu, D.~Abusch-Magder, U.~Meirav, and
  M.~A. Kastner.
\newblock {\em Nature}, 391:156, 1998.

\bibitem{thomas00}
K.~J. Thomas, J.~T. Nicholls, M.~Pepper, W.~R. Tribe, M.~Y. Simmons, and D.~A.
  Ritchie.
\newblock {\em Phys.\ Rev.\ B}, 61:13365, 2000.

\bibitem{appleyard}
N.~J. Appleyard, J.~T. Nicholls, M.~Pepper, W.~R. Tribe, M.~Y. Simmons, and
  D.~A. Ritchie.
\newblock {\em Phys.\ Rev.\ B}, 62:16275, 2000.

\bibitem{roche}
P.~Roche, J.~Segala, D.~C. Glattli, J.~T. Nicholls, M.~Pepper, A.~C. Graham,
  K.~J. Thomas, M.~Y. Simmons, and D.~A. Ritchie.
\newblock {\em Phys. Rev. Lett.}, 93:116602, 2004.

\end{thebibliography}

\end{document}